\documentclass[runningheads]{llncs}
\usepackage{hyperref}

\usepackage{threeparttable}

\usepackage{adjustbox}

\usepackage{amssymb}

\usepackage{booktabs}

\usepackage{amsmath} 

\usepackage[T1]{fontenc}
\usepackage{graphicx}
\begin{document}
\title{TransUNext: towards a more advanced U-shaped
framework for automatic vessel segmentation in the
fundus image}
%
%
%
%
\author{Xiang Li\inst{1} \and
Mingsi Liu\inst{1} \and
Lixin Duan\inst{1}}
\authorrunning{X. Li et al.}
%
\institute{
Shenzhen Institute for Advanced Study, University of Electronic Science and Technology of China, Shenzhen, 518110, China
}

\maketitle              

\begin{abstract}
\textbf{Purpose}: Automatic and accurate segmentation of fundus vessel images has become an essential prerequisite for computer-aided diagnosis of ophthalmic diseases such as diabetes mellitus. The task of high-precision retinal vessel segmentation still faces difficulties due to the low contrast between the branch ends of retinal vessels and the background, the long and thin vessel span, and the variable morphology of the optic disc and optic cup in fundus vessel images.

\textbf{Methods}: We propose a more advanced U-shaped architecture for a hybrid Transformer and CNN: TransUNext, which integrates an Efficient Self-attention Mechanism into the encoder and decoder of U-Net to capture both local features and global dependencies with minimal computational overhead. Meanwhile, the Global Multi-Scale Fusion (GMSF) module is further introduced to upgrade skip-connections, fuse high-level semantic and low-level detailed information, and eliminate high- and low-level semantic differences. Inspired by ConvNeXt, TransNeXt Block is designed to optimize the computational complexity of each base block in U-Net and avoid the information loss caused by the compressed dimension when the information is converted between the feature spaces of different dimensions.

\textbf{Results}: We evaluated the proposed method on four public datasets DRIVE, STARE, CHASE-DB1, and HRF. In the experimental results, the AUC (area under the ROC curve) values were 0.9867, 0.9869, 0.9910, and 0.9887, which exceeded the other state-of-the-art (SOTA) methods, respectively. In addition, evaluation metrics such as CAL (connectivity-area-length) were used in the ablation study to quantify the segmentation of coarse and fine vessels.

\textbf{Conclusion}: The quantitative and qualitative results validate the superior performance of the proposed method for retinal vessel segmentation, and the robustness in challenging situations, such as lesions or fine vessels in the optic disc region, is well demonstrated. The Source code is available at \href{https://github.com/lixiang007666/TransUNext}{https://github.com/lixiang007666/TransUNext}.

\begin{flushleft}
\textbf{Keywords}: Fundus image, vessel segmentation, TransNeXt Block, Global Multi-Scale Fusion
\end{flushleft}

\end{abstract}

\section{Introduction}
The essence of automatic segmentation of fundus vessel images is to dichotomize the vascular pixels in the image with the surrounding pixels. In clinical applications, fundus vessel images are more complex. They require experienced professionals to complete segmentation manually. Not only are they subjective and inefficient, but with the explosion of fundus image data, implementing computer-aided automatic segmentation of vessel networks in fundus images has crucial clinical value \cite{Luo_2020}.

Currently, automatic segmentation methods for fundus vessel images are divided into two main categories: one is based on unsupervised methods, and the other is based on supervised methods, including machine learning and deep learning strategies.

Based on unsupervised machine learning methods, Chaudhuri et al. \cite{34715} successfully implemented blood vessel segmentation of fundus images using a two-dimensional Gaussian matched filter. Afterward, some segmentation methods based on vessel morphology and particular pixels appeared. For example, Yang et al. \cite{10.5555/2975865.2975874} proposed a morphological processing method, which first enhances vessel features, suppresses background information, and then applies fuzzy clustering to achieve vessel segmentation; Zhao et al. \cite{7055281} also proposed a segmentation method based on a deformable model, which uses regional information of different vessel types to achieve segmentation; Li et al. \cite{LI20127600} optimized the matched filtering method and applied it to the vessel segmentation task. The segmentation method based on unsupervised learning is fast, but the segmentation result is rough and has low accuracy.

The supervised machine learning segmentation method better extracts vessel feature information by strengthening the training model through manually labeled images. Staal et al. \cite{1282003} and Soares et al. \cite{1677727} used a two-dimensional filter to extract the overall features of the retinal image and then used naive Bayes to classify the retinal background and blood vessels. Ricci et al. \cite{4336179} first extracted the green channel of the fundus image during image preprocessing and then used SVM to segment according to the difference in vessel width. Fraz et al. \cite{6224174} proposed combining AdaBoost and Bagging models, integrating the results of complex feature extraction and the results of binary classification models, and using the supervised method to segment retinal vessel images automatically. Although the accuracy of the supervised machine learning method has been improved, because the algorithm itself cannot adapt to the shape, scale, and geometric transformation of blood vessels, there are still problems, such as low accuracy and low robustness when segmenting small vessels and vessel intersections, and it is difficult to provide an objective basis for clinical diagnosis.

With the advent of CNN, the semantic segmentation method based on deep learning can accurately predict vessel and non-vessel pixels and provide descriptions of vessel scale, shape, multiple curvature, and other information. In medical image semantic segmentation methods, U-Net \cite{10.1007/978-3-319-24574-4_28} is considered a very successful network, consisting of convolutional encoding and decoding units. It can use a few samples to complete training to perform segmentation tasks better. Its derivative works \cite{10.1007/978-3-319-46723-8_17,7420682,10.1007/978-3-030-00934-2_14,8341481} have also achieved advanced retinal blood vessel segmentation results. In order to further improve the accuracy of retinal blood vessel segmentation, Wang B et al. \cite{10.1007/978-3-030-32239-7_10} proposed a variant of U-Net with dual encoders to capture richer context features. Li et al. \cite{Wang2021} also proposed an improved end-to-end network of U-Net. This framework uses technologies including compression and excitation (SE) module, residual module, and circular structure and introduces enhanced super-resolution generative adaptive networks (ESRGANS) \cite{Wang_2018_ECCV_Workshops} and improved data enhancement methods to achieve retinal blood vessel segmentation. Wang B et al. \cite{9098742} also provided a supervision framework for retinal vessel segmentation from thick to thin.

Nevertheless, CNN-based approaches cannot model long-range dependencies due to inherent inductive biases such as locality and translational equivalence. Thus Transformer \cite{NIPS2017_3f5ee243}, which relies purely on attention mechanisms to build global dependencies without any convolutional operations, has emerged as an alternative architecture that provides better performance than CNNs in computer vision (CV) under pre-training conditions on large-scale datasets. Vision Transformer (Vit) \cite{2010.11929} revolutionized the CV field by segmenting images into a sequence of tokens and modeling their global relationships with stacked Transformer blocks. Swin Transformer \cite{Liu_2021_ICCV} can produce hierarchical features in a movable window with low computational complexity representation, achieving state-of-the-art performance in various CV tasks. However, the size of medical image datasets is much smaller than the pre-trained datasets in the above work (e.g., ImageNet-21k and JFT-300M). As a result, the Transformer produces unsatisfactory performance in medical image segmentation. Therefore, many hybrid structures combining CNN and Transformer have emerged, which have both advantages and are gradually becoming a compromise solution for medical image segmentation without needing pre-training on large datasets.

\begin{figure}
  \centering
  \includegraphics[width=1\textwidth]{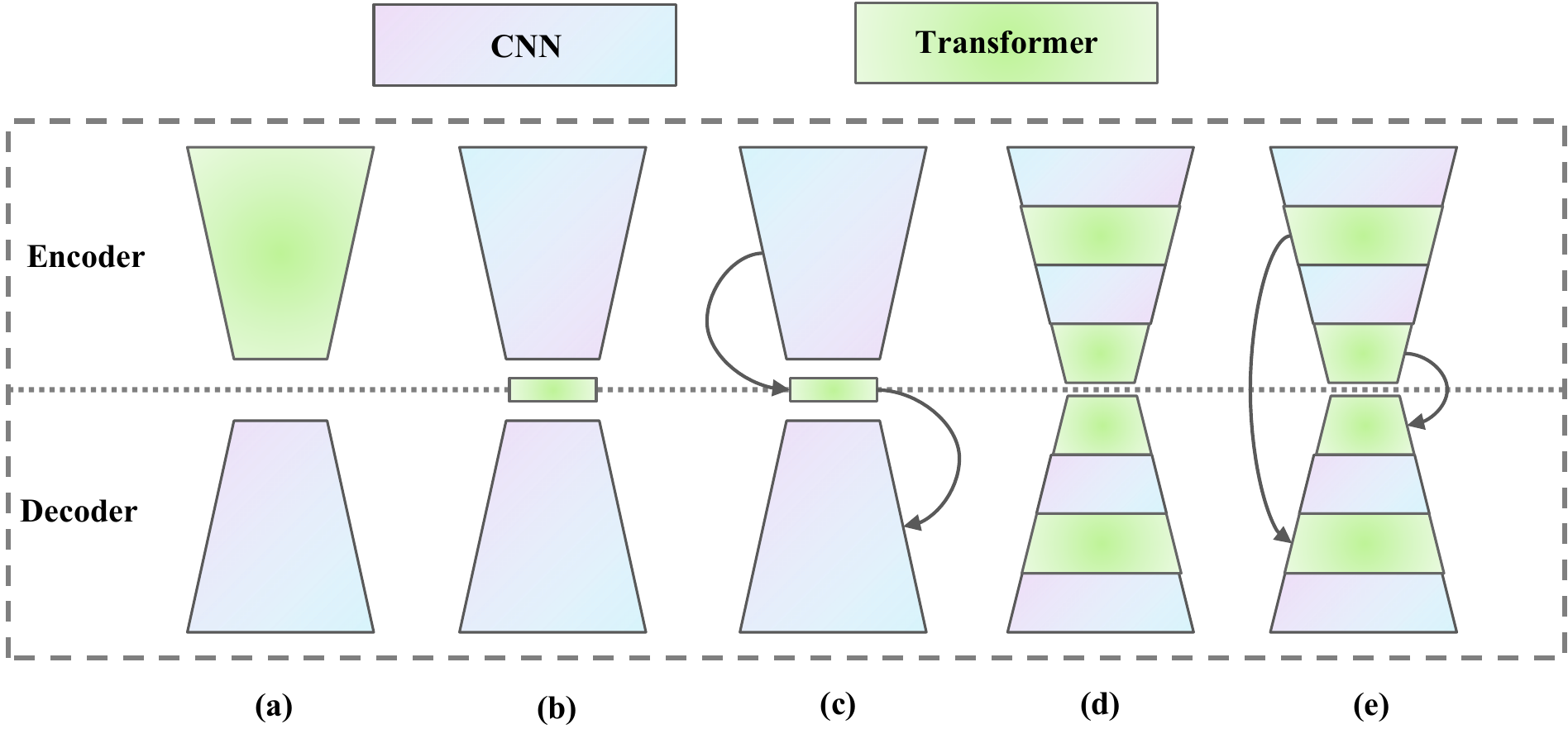}
  \caption{Comparison of several hybrid architectures of Transformer and CNN. (Color figure online)}
  \label{fig:img1}
\end{figure}

We summarize several popular hybrid architectures based on Transformer and CNN in medical image segmentation. These hybrid architectures add the Transformer to a CNN-based backbone model or replace some architecture components. For example, UNETR \cite{Hatamizadeh_2022_WACV} uses an encoder and decoder architecture where the encoder is composed of a cascaded block built with a pure Transformer, and the decoder is a stacked convolutional layer, see Fig. \ref{fig:img1}(a).TransBTS \cite{10.1007/978-3-030-87193-2_11} and TransUNet \cite{2102.04306} introduce a relationship between the encoder and decoder composed of a CNN a Transformer, see Fig. \ref{fig:img1}(b). coTr \cite{10.1007/978-3-030-87199-4_16} bridges all stages from the encoder to the decoder through the Transformer, not only the adjacent stages, allowing to exploit the global dependencies at multiple scales, see Fig. \ref{fig:img1}(c). Furthermore, nn-Former \cite{2109.03201,2105.05537} interweaves Transformer and convolutional blocks into a hybrid model where convolution encodes precise spatial information while Transformer captures global context information, see Fig. \ref{fig:img1}(d). As seen from Fig. \ref{fig:img1}, these architectures implement a serial combination of Transformer and CNN from a macroscopic perspective. However, in these combinations, the convolutional and self-attention mechanisms cannot be applied throughout the network structure, making it challenging to model local and global features efficiently.

In this paper, we propose a more advanced U-shaped architecture of a hybrid Transformer for fundus vessel image segmentation (TransUNext), the pattern of which is shown in Fig. \ref{fig:img1}(e), and TransUNext combines the advantages of both Fig. \ref{fig:img1}(c) and Fig. \ref{fig:img1}(d). It can take full advantage of the global dependencies at multi-scale, fuse high-level semantic and low-level detail information, and efficiently aggregate local and global features. In addition, inspired by ConvNeXt \cite{Liu_2022_CVPR}, we also optimize each base convolutional block in the U-shaped architecture to maintain excellent segmentation capability even at low computational complexity.

We found that TransUNext can further improve the accuracy of vessel segmentation with the following main contributions:

\begin{itemize}
    \item TransNeXt Block is proposed as the base block in the U-shaped architecture. This base block avoids the information loss caused by the compressed dimension when the information is converted between the feature spaces of different dimensions and uses self-attention to capture global dependencies. At the same time, it avoids large-scale pre-training of the Transformer and has a more negligible computational overhead.
    \item Upgrading the skip-connection of U-Net using Global Multi-Scale Fusion (GMSF) structure and propagating information through all-to-all attention in all tokens at each scale achieves semantic and spatial global multi-scale fusion, which can make full use of high-level semantic and low-level detail information.
    \item The proposed method was evaluated on four public datasets, DRIVE, STARE, CHASE-DB1, and HRF, where the images of HRF are of high resolution. In addition, in addition to some general evaluation metrics, such as AUC, we used CAL for the specificity of the retinal vessel segmentation task \cite{6019055}.
\end{itemize}

The rest of this paper is organized as follows: Section II introduces the proposed method in detail. Section III describes the experimental implementation and illustrates the experimental results. Section IV gives the conclusions of this paper

\section{Proposed method} 
\subsection{Dataset Pre-processing}

\begin{figure}
  \centering
  \includegraphics[width=1\textwidth]{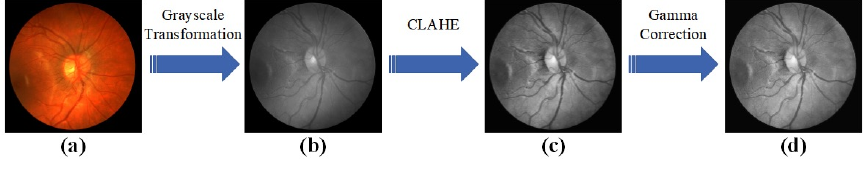}
  \caption{The flowchart of data preprocessing.}
  \label{fig:img2}
\end{figure}

Many samples have poor contrast and high noise in the fundus vessel image dataset. Therefore, proper preprocessing is crucial for subsequent training. This paper uses four preprocessing methods, including Grayscale Transformation, Data Normalization, Contrast Limited Adaptive Histogram Equalization (CLAHE), and Gamma Correction \cite{6019055,8662594,10.1145/3348416.3348425}, to process each original fundus vessel image. Fig. \ref{fig:img2} provides a schematic diagram of the phased processing results of the original color retinal images after Grayscale Transformation, CLAHE, and Gamma Correction. As can be seen from the figure, the preprocessed images have clear texture, prominent edges, and enhanced detail information.

\begin{figure}
  \centering
  \includegraphics[width=1\textwidth]{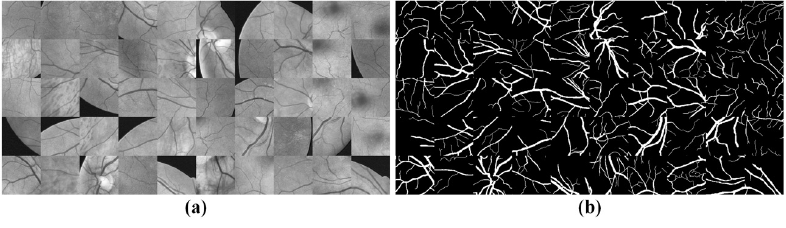}
  \caption{Illustration of random cropping. (a) patches from the original image (b) patches of ground truth.}
  \label{fig:img3}
\end{figure}

Because of the relatively small amount of data in the fundus vessel image dataset, a new augmentation was performed to reduce the effects of overfitting. The retinal vessel image in the used dataset is a circular region, so rotating the image randomly by a fixed angle can simulate different acquisition environments without changing the structure of the image itself. In addition, 15,000 randomly extracted patches of size 128 in each training image of the DRIVE, STARE, CHASE-DB1, and HRF datasets were extracted by random cropping, and the corresponding ground truth was processed identically. The amplified image data are shown in Fig. \ref{fig:img3}.

Unlike the cropping method in the training phase, in the testing phase, each image block needs to be re-stitched into a complete image and then binarized to obtain the segmentation result. All patches must be stitched to restore their resolution to the level of the original fundus image. However, the time and space complexity of stitching based on the index is extremely high if random cropping is used. To avoid this problem, we use overlapping cropping in the testing phase. The step size was set to 12 based on workstation performance trade-offs.

\subsection{Overall Architecture}

\begin{figure}
  \centering
  \includegraphics[width=1\textwidth]{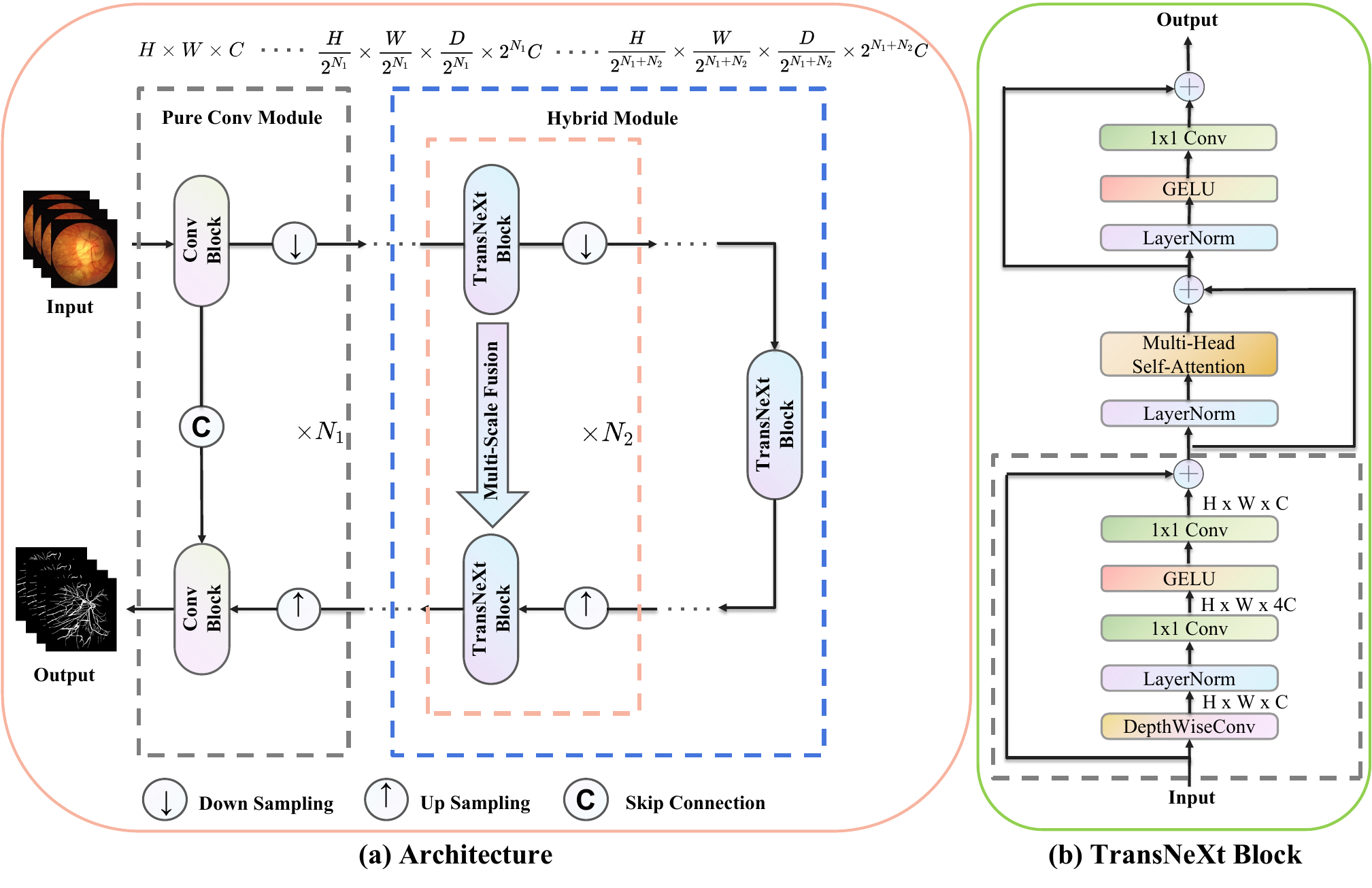}
  \caption{(a) The architecture of TransUNext; (b) Hybird block consisting of Transformer and ConvNeXt (TransNeXt Block).}
  \label{fig:img4}
\end{figure}

An overview of the TransUNext architecture is illustrated in Fig. \ref{fig:img4}(a). TransUNext follows the design of a U-shaped network encoder and decoder, mainly consisting of pure convolutional modules and TransNeXt Blocks. While our original intention was to build a fully hybrid architecture consisting of Transformer and ConvNeXt, in our implementation, cascaded pure convolutional modules and downsampling operations are introduced to gradually extract low-level features with high resolution and obtain delicate spatial information. Similarly, these pure convolutional modules are deployed in the same stage of the decoder to recover the original image dimension by upsampling. Note that we do not apply the Transformer on the original resolution, as adding the Transformer module in the very shallow layers of the network does not help in experiments but introduces additional computation. A possible reason is that the shallow layers of the network focus more on detailed textures, where gathering global context may not be informative \cite{10.1007/978-3-031-16443-9_23}.

It is worth noting that TransUNext does not use a simple skip connection between the encoder and decoder. Instead, Global Multi-Scale Fusion (GMSF), which propagates information through all-to-all attention in all tokens at each scale, enables semantic and spatial global multi-scale fusion, which can fully use high-level semantic and low-level detail information.

Given an input volume $x \in \mathbb{R}^{H \times W \times D}$, where $H$, $W$ and $D$ denote the height, width, and depth, respectively, we first utilize several pure convolution modules to obtain feature maps $f \in \mathbb{R} \frac{H}{2^{N_1}} \times \frac{W}{2^{N_1}} \times \frac{D}{2^{N_1}} \times 2^{N_1} C$, where $N_1$ and $C$ denote the number of modules and base channels, respectively. Afterwards, hybrid modules consisting of Transformer and ConvNeXt were applied to model the hierarchical representation from local and global feature. The procedure is repeated $N_2$ times with $\frac{H}{2^{N_1+N_2}} \times \frac{W}{2^{N_1+N_2}} \times \frac{D}{2^{N_1+N_2}}$ as the output resolutions and $2^{N_1+N_2} C$ as the channel number. The architecture of TransUNext is straightforward and changeable, where the number of each module can be adjusted according to medical image segmentation tasks, i.e., $N_1$ and $N_2$.

\subsection{TransNeXt Block}

The hybrid modules are deployed in the deep stages of TransUNext, where the TransNeXt Block, as its heart, achieves aggregation of local and global representations by ConvNeXt and Transformer. Fig. \ref{fig:img4}(b) shows the structure of the TransNeXt Block from bottom to top, and the essential components include DepthWise (DW) Conv, LayerNorm, 1×1 Conv, GELU, multi-head self-attention (MHSA), and three residual connections. We use this structure as the base block of the U-shaped architecture, which captures local features and avoids the information loss caused by compressing dimensions when the input is transformed between feature spaces of different dimensions. In addition, Transformer's self-attention mechanism can efficiently capture global dependencies.

\textbf{Revisiting ConvNeXt.} Liu et al. have proposed a convolutional neural network model for the 2020s ConvNeXt network. The author uses the ideas of Swin-transformer, ResNeXt \cite{Xie_2017_CVPR}, and MobileNet \cite{1704.04861} to change the calculation amount in different stages at the macro level, convolution with large-scale convolution kernel, group convolution, etc. In the micro design, the RELU activation function is replaced by the GELU activation function, fewer activation functions and normalization layers are used, and LN replaces BN. The network outperforms the Swin-t model in many classification and recognition tasks to achieve the best performance.

Specifically, as shown in the header of Fig. \ref{fig:img4}(b), the input first passes through the DepthWise Conv with kernel size=7, stride=1 and padding=3, aiming to get better results with less number of parameters compared to the standard base convolution block. At the channel level, its size changes from C to 4C and then decreases to C. Both the ConvNeXt block adopts the form of small, large, and small dimensions to avoid the information loss caused by the compressed dimension when the information is converted between the feature spaces of different dimensions.

\textbf{Efficient Self-attention Mechanism}. As can be found in Fig. \ref{fig:img4}(b), the TransNeXt Block is built based on the core MHSA module. This module allows the model to infer attention from different representation subspaces jointly. The results from multiple heads are concatenated and then transformed with a feed-forward network. In this paper, we used four heads; you can see more detailed multi-head dimensions in the code.

\begin{figure}
  \centering
  \includegraphics[width=1\textwidth]{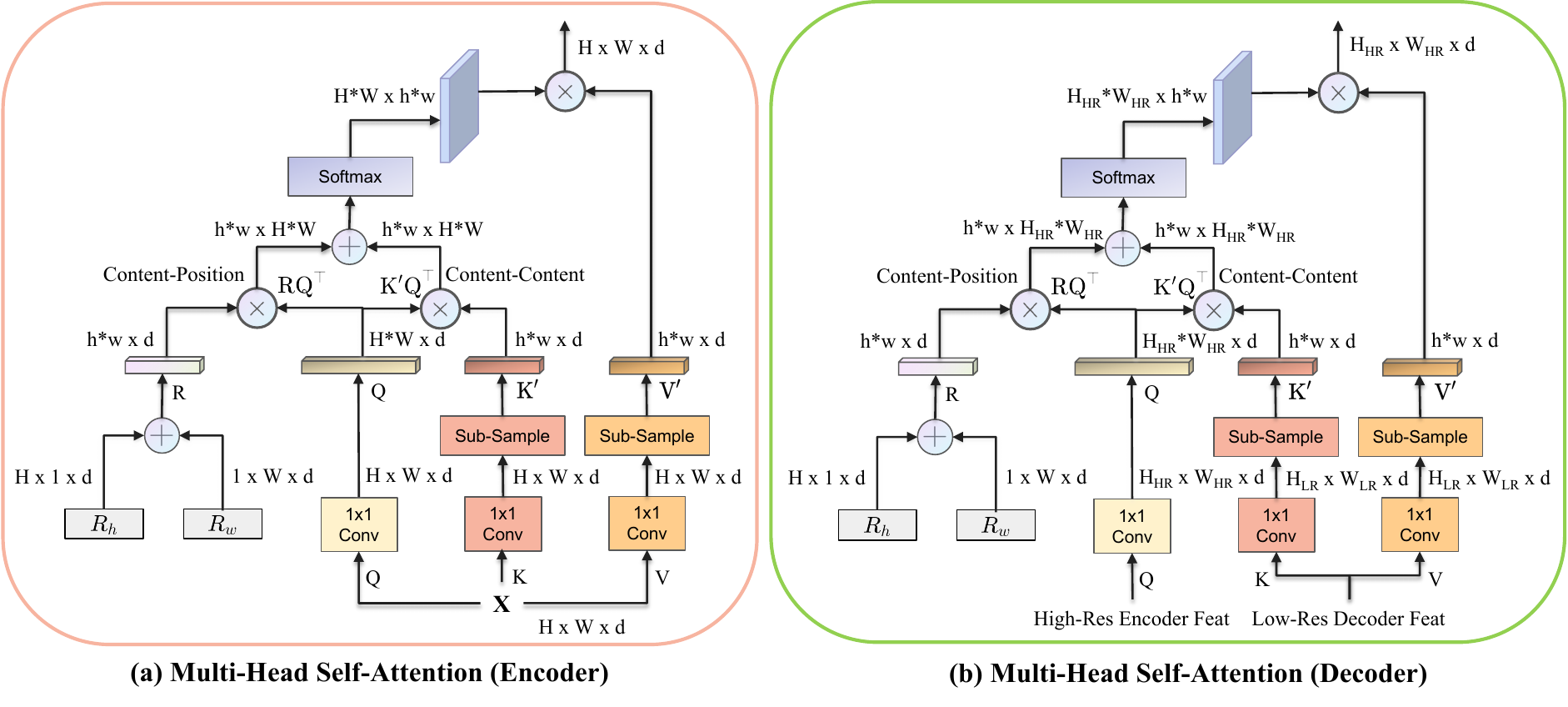}
  \caption{Efficient multi-head self-attention (MHSA). (a) The MHSA used in the Transformer encoder. (b) The MHSA used in the Transformer decoder. They share similar concepts, but (b) takes two inputs, including the high-resolution features from GMSF of the encoder, and the low-resolution features from the decoder.}
  \label{fig:img5}
\end{figure}

As shown in Fig. \ref{fig:img5}. Three 1×1 convolutions are used to project X to query, key, value embeddings: $\mathbf{Q}, \mathbf{K}, \mathbf{V} \in \mathbb{R}^{d \times H \times W}$, where $d$ is the dimension of embedding in each head. The $\mathbf{Q}, \mathbf{K}, \mathbf{V}$ is then flatten and transposed into sequences with size $n \times d$, where $n=H W$. Self-attention mechanism is implemented by the dot product of vectors and softmax. However, images are highly structured data, and most pixels in a high-resolution feature map have similar features in the neighborhood except for the boundary regions. Therefore, pairwise attention computation in the whole image is unnecessary \cite{2006.04768}.

So we use an Efficient Self-attention Mechanism corresponding to the Sub-Sample component in Fig. \ref{fig:img5}. The main idea is to use two projections for keys and values, and the processed self-attention is computed as follows:

\begin{equation}
\text { Attention }\left(\mathbf{Q}, \mathbf{K}^{\prime}, \mathbf{V}^{\prime}\right)=\operatorname{softmax}\left(\frac{\mathbf{K}^{\prime} \mathbf{Q}^{\top}}{\sqrt{d}}\right) \mathbf{V}^{\prime}
\end{equation}
where $\mathbf{K}, \mathbf{V} \in \mathbb{R}^{n \times d}$ into low-dimensional embedding: $\mathbf{K}^{\prime}, \mathbf{V}^{\prime} \in \mathbb{R}^{k \times d}$, and $k=h w \ll n$, $h$ and $w$ are the reduced size of feature map after sub-sampling. Among them, the shape of softmax $\left(\frac{\mathbf{K}^{\prime} \mathbf{Q}^{\top}}{\sqrt{d}}\right)$ is $n \times k$, and the shape of $\mathbf{V}^{\prime}$ is $k \times d$.

By doing so, the computational complexity is reduced to $O(n k d)$, instead of the $O\left(n^2 d\right)$ complexity of a simple dot product. Notably, the projection to low-dimensional embedding can be any down-sampling operations, such as strided convolution, or average pooling/max pooling. The advantage of this is that, for example, with a high-resolution fundus image dataset like HRF, n is much larger than d because of the high-resolution of its feature maps. Thus the sequence length dominates the self-attention computation and makes it infeasible to apply self-attention in high-resolution feature maps. In our implementation, we still use 1×1 convolution followed by a simple interpolation to down-sample the feature map.

\subsection{Global Multi-Scale Fusion}

\begin{figure}
  \centering
  \includegraphics[width=1\textwidth]{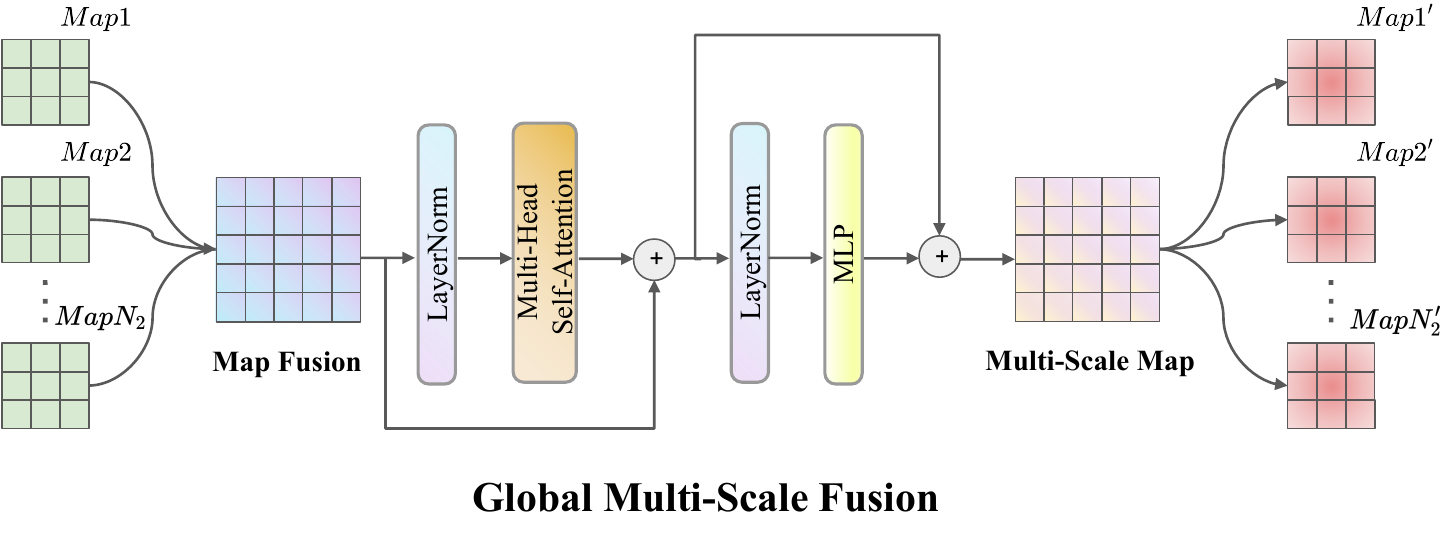}
  \caption{The proposed Global Multi-Scale Fusion(GMSF) module.}
  \label{fig:img6}
\end{figure}

Fig. \ref{fig:img6} shows a spatially and semantically Global Multi-Scale Fusion (GMSF) module that integrates information from multi-scale semantic graphs with a small computational overhead. In the decoder, TransUNext progressively recovers the resolution through TransNeXt Block and upsampling layers and combines the high-resolution feature maps from the encoder through GMSF in place of the traditional U-network skip connections to finally output the segmentation maps.

Global Multi-Scale Fusion plays a crucial role in dense prediction tasks, combining high-level semantic and low-level detail information, while the feature maps output in TransNeXt Block are naturally used in the Global Multi-Scale Fusion module. First, accepting 2D feature maps from multiple scales: \\$M a p_1, M a p_2, \ldots, M a p_n$, Map Fusion contains feature maps from all scales, which are obtained by concatting and reshaping the feature maps from multiple inputs. Then it is sent to the Transformer block for multi-scale semantic fusion. Finally, the fused Multi-Scale Map is chunked and reshaped into a 2D feature map. In contrast to the skip connection, the proposed approach propagates information to all tokens at each scale by all-to-all attention to form a semantic and spatial Global Multi-Scale Fusion. i.e., the "skip connection" now has an attention mechanism that can eliminate high and low-level semantic differences and improve feature reuse.

\subsection{Evaluation metrics}

To evaluate the performance of the proposed method, several metrics are used in this paper, including Accuracy (\textit{Acc}), Specificity (\textit{SP}), Sensitivity (\textit{SE}), \textit{Precision}, \textit{F1-Score}, \textit{AUC} (area under the \textit{ROC} curve) and \textit{CAL} (connectivity-area-length) \cite{6019055,10.1007/978-3-031-16434-7_51}.

\textit{Acc}: It measures the proportion of correctly classified pixels, a representative evaluation metric for vessel segmentation tasks.

\begin{equation}
A c c=\frac{T P+T N}{T P+F P+F N+T N}
\end{equation}

\textit{SP}: It refers to the proportion of correctly classified non-vessel pixels to actual non-vessel pixels.

\begin{equation}
S P=\frac{T N}{T N+F P}
\end{equation}

\textit{SE}: It is also known as recall rate (\textit{Recall}), which refers to the proportion of correctly classified blood vessel pixels to actual blood vessel pixels.

\begin{equation}
\mathrm{SE}=\frac{T P}{T P+F N}
\end{equation}

\textit{Precision}: It refers to the proportion of correctly classified vessel pixels to all vessel pixels.

\begin{equation}
\text { Precision }=\frac{T P}{T P+F P}
\end{equation}

\textit{F1-Score}: It measures the binary classification model, which considers the model's precision and recall rate. The \textit{F1-score} can be seen as the harmonic mean of the precision and recall, showing satisfactory results when \textit{SE} and \textit{Precision} values are high.

\begin{equation}
F_1-\text { Score }=2 * \frac{\text { Precision } * S E}{\text { Precision }+S E}=\frac{T P}{T P+\frac{F P+F N}{2}}
\end{equation}

In addition, we depicted the receiver operating characteristic curve (\textit{ROC}), which was generated with \textit{TP} as the ordinate and \textit{FP} as the abscissa. We also provided the Area under the \textit{ROC} curve (\textit{AUC}), which considers the \textit{SE} and \textit{SP} under different thresholds, and is suitable for measuring retinal vessel segmentation.

Also, some metrics have been specially designed for vessel segmentation and widely used in previous works. For example, a set of metrics was proposed by Gegundez et al. [36] to evaluate the connectivity (\textit{C}), overlapping area (\textit{A}), and consistency of vessel length (\textit{L}) of predicted vessels. The overall metric (\textit{F}) was defined as

\begin{equation}
F(C, A, L)=C \times A \times L .
\end{equation}
In doing so, the segmentation of coarse and fine vessels can be quantified more equally.

\section{Experiments} 

\subsection{Datasets}
\begin{figure}
  \centering
  \includegraphics[width=1\textwidth]{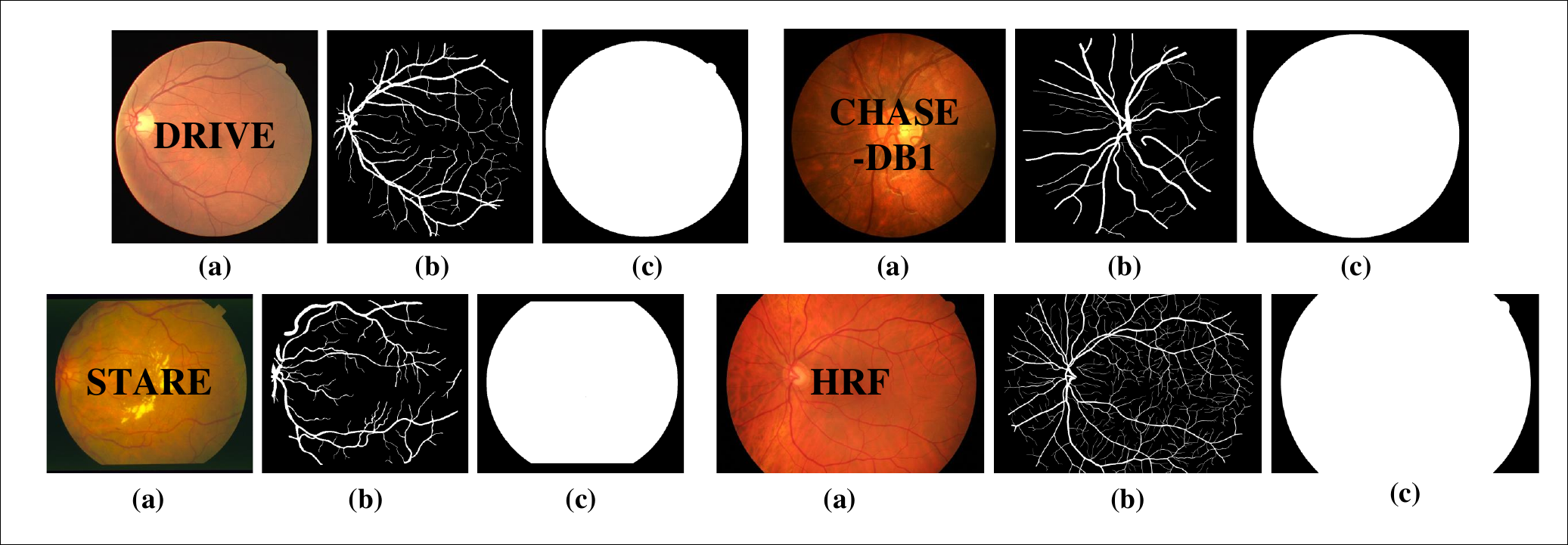}
  \caption{Fundus images from four different datasets. (a) Original fundus image (b) ground truth (c) mask.}
  \label{fig:img7}
\end{figure}

We evaluated our proposed method on four public datasets of fundus images, including DRIVE, STARE, CHASE-DB1, and HRF. Fig. \ref{fig:img7} shows some typical cases from the four datasets.

\begin{enumerate}
    \item DRIVE Dataset: The DRIVE dataset \\(\url{https://http://drive.grand-challenge.org/DRIVE}) contains 40 fundus retinal color images, seven of which are from patients with early diabetic retinopathy, with a resolution of 565 × 584 and stored in JPEG format. The original dataset uses 20 images for training and 20 for testing with masks, and two experts manually annotated the dataset. In this paper, we divide the dataset into a training set, a validation set, and a test set according to the ratio of 18:2:20.
    \item STARE Dataset: The STARE dataset \\(\url{http://cecas.clemson.edu/~ahoover/stare/probing}) provides 20 fundus color images with a resolution of 605 × 700. We use 15 of these images for training and five for testing. The original dataset is not divided into a validation set like the DRIVE dataset. Thus, we choose 10\% of the training data for validation. The STARE dataset also provides annotated images of two experts.
    \item CHASE-DB1 Dataset: The CHASE-DB1 dataset \\(\url{https://blogs.kingston.ac.uk/retinal/chasedb1}) contains 28 color retinal images with a resolution of 996 × 960. It was taken from the left and right eyes of 14 children. We used 21 of these images for training and seven for testing.
    \item HRF Dataset: The are 45 fundus images in the HRF dataset \\(\url{https://www5.cs.fau.de/fileadmin/research/datasets/fundus-images}) with a resolution of 3504 × 2336. Thirty-eight images from each group of healthy children, diabetic retinopathy, and glaucoma patients are taken as the training set, and the other seven images are taken as the test set.
\end{enumerate}

The detailed division of the four datasets is shown in Table \ref{table1}.

\begin{table}[!htbp]
  \centering  
  \caption{The division method of datasets for experiment.} 
  \label{table1} 
    \begin{tabular}{cccc}
    \toprule
    \textbf{Database}  & \textbf{Training}  & \textbf{Validation} & \textbf{Test}      \\
    \midrule
    DRIVE     & 18 images & 2 images   & 20 images \\
    STARE     & 13 images & 2 images   & 5 images  \\
    CHASE-DB1 & 19 images & 2 images   & 7 images  \\
    HRF       & 34 images & 4 images   & 7 images  \\
    \bottomrule
    \end{tabular}
    \begin{tablenotes}
     \item[1] \textbf{Note}: Set up division based on experience and datasets requirements.
   \end{tablenotes}
\end{table}

\subsection{Implementation details}

We implemented TransUNext under torch-1.9.0-cu11.1-cudnn8 and conducted experiments with a single GeForce RTX 2080Ti. 

In the training process, we set the maximum number of training epochs to 25, batch size to 8, and initial learning rate to 0.0005. A binary cross-entropy loss (BCE) is used as the objective function to supervise the model's training process. The optimizer that performs gradient updates uses Adam. The DRIVE, STARE, CHASE-DB1, and HRF datasets follow the same data augmentation strategy: randomly extract image patches with a resolution of 128 × 128 from the preprocessed images. In TransUNext, we empirically set the hyper-parameters [$N_1$,$N_2$] to [1,3].

Unlike the training phase, we perform overlap cropping in the original image with a fixed step size. Since these patches have overlapping areas (I. e., each pixel appears multiple times in different patches), we averaged the probability value of each pixel belonging to retinal blood vessels and set the threshold to obtain a binarized prediction map. In addition, we use an early stopping mechanism to prevent the occurrence of overfitting.

\subsection{Results}
\subsubsection{Ablation study}

We conducted an ablation study on the DRIVE dataset to verify each module's contribution to the entire model's performance. As can be seen from Table \ref{table2}, the AUC, C, A, and L of TransUNext reached 0.9867, 99.85\%, 93.58\%, and 91.28\%, respectively. Compared with the baseline model in the DRIVE dataset, the performance of the final model is improved by 0.8\%, 1.05\%, 1.11\%, and 1.09\%, respectively.

Although the FLOPs and Params of TransUNext are higher than those of Baseline, the increase in memory consumption and storage space is acceptable, considering the improvement in its segmentation accuracy. Comparing the data of the two rows of \textit{Baseline w/CNN and Trans block} and \textit{Baseline + all (our TransUNext)}, the structure of the former follows the design of Fig. \ref{fig:img1}(e); it is found that our proposed TransNeXt Block, which can optimize the computational complexity of each base block in U-Net, has a higher performance with a lower number of parameters.

\begin{table}[!htbp]
  \centering  
  \caption{Performance of proposed method tested on four datasets.} 
  \label{table2} 
  \begin{adjustbox}{center}
    \begin{tabular}{ccccccc}
    \toprule
    \textbf{Methods}                         & \textbf{AUC}    & \textbf{C(\%)} & \textbf{A(\%)} & \textbf{L(\%)} & \textbf{FLOPs/G} & \textbf{Params/M} \\
    \midrule
    Baseline w/pure CNN block       & 0.9787 & 98.8  & 92.47 & 90.19 & \textbf{12.97}   & \textbf{7.07}     \\
    Baseline w/CNN and Trans block  & 0.9818 & 98.89 & 93.51 & 90.89 & 21.51   & 9.53     \\
    \midrule
    Baseline + GMSF                 & 0.983  & 99.78 & 93.57 & 91.08 & 21.78   & 9.74     \\
    Baseline + TransNeXt            & 0.986  & 99.84 & 93.48 & 91.15 & 18.32   & 8.34     \\
    Baseline + all (our TransUNext) & \textbf{0.9867} & \textbf{99.85} & \textbf{93.58} & \textbf{91.28} & 18.51   & 8.55     \\
    \bottomrule
    \end{tabular}
    \end{adjustbox}
    \begin{tablenotes}
     \item[1] \textbf{Note}: For each metric, the bold value indicates that column's best result. Baseline is a standard U-Net.
   \end{tablenotes}
\end{table}

\begin{figure}
  \centering
  \includegraphics[width=0.75\textwidth]{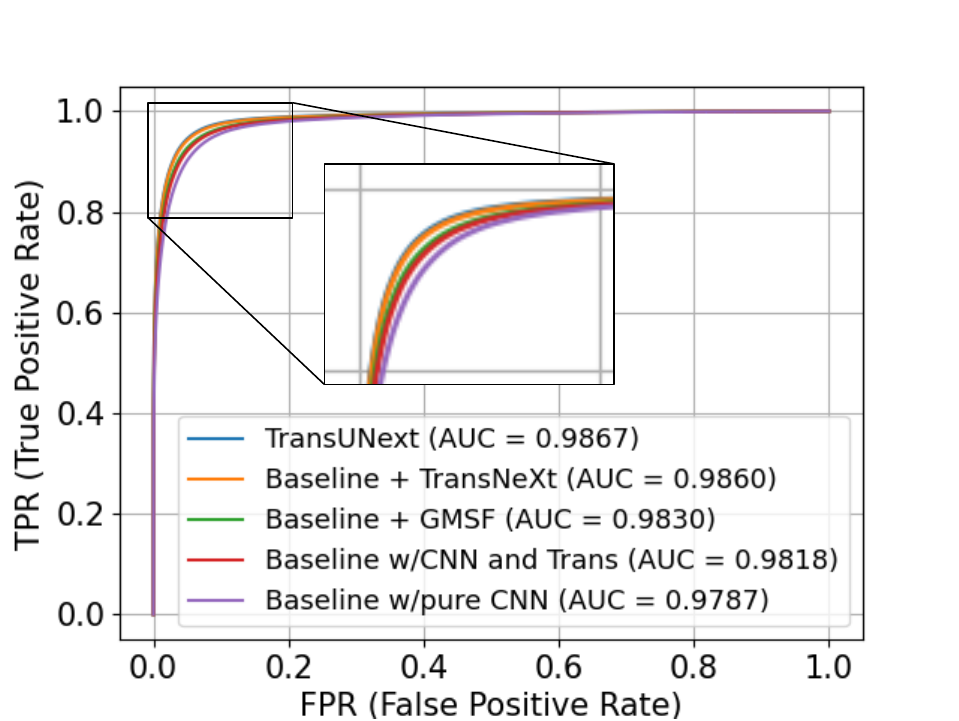}
  \caption{ROC curves of ablation study on the DRIVE dataset.}
  \label{fig:img8}
\end{figure}

To better evaluate the whole ablation study, Fig. \ref{fig:img8} shows the ROC curves of TransUNext on the test dataset of DRIVE. After visualization, it can also be found that both GMSF and TransNeXt Block are designed to enhance the segmentation performance with AUC values of 0.9830 and 0.9860, respectively.

\subsubsection{Results on four datasets}

\begin{figure}
  \centering
  \includegraphics[width=1\textwidth]{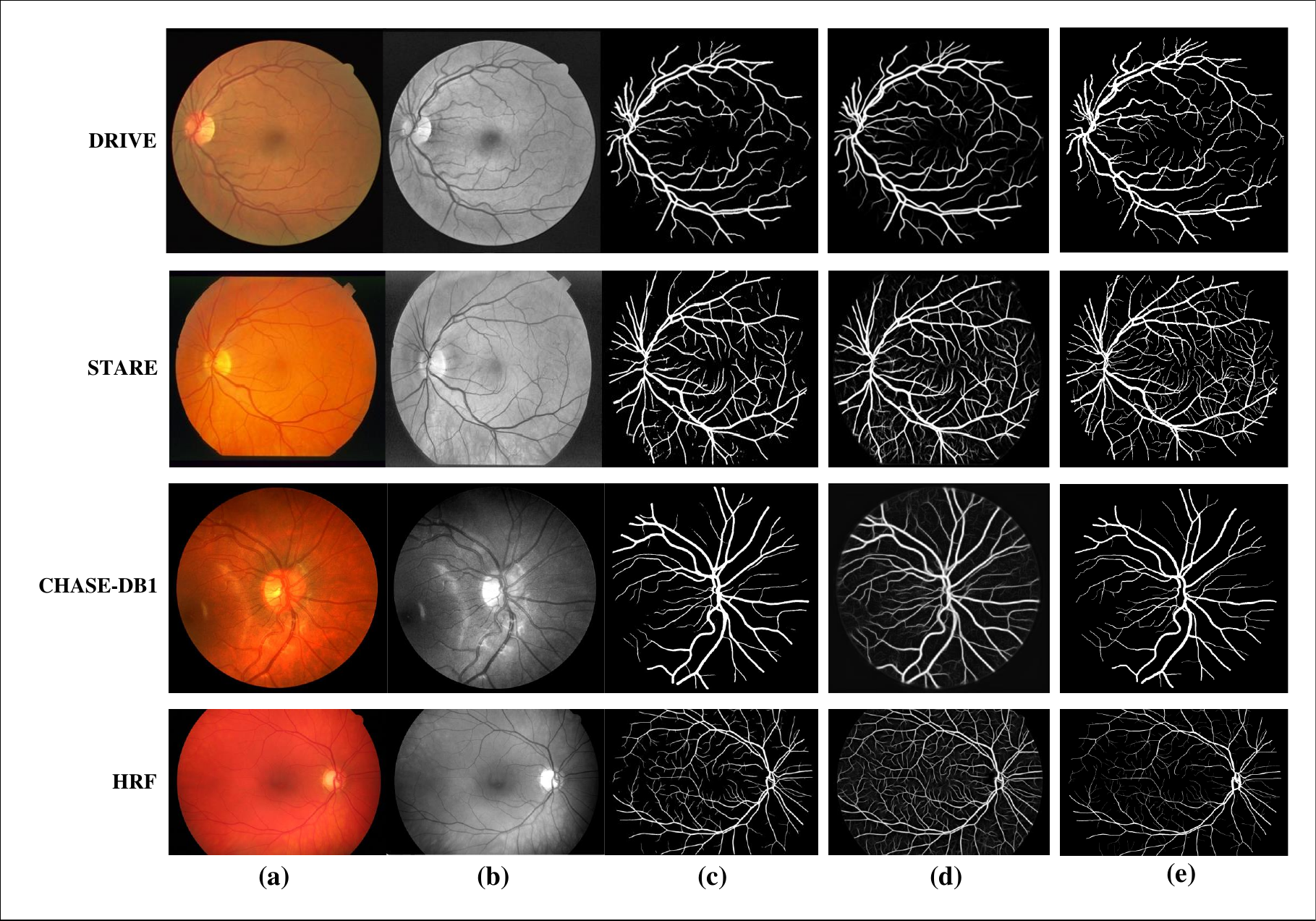}
  \caption{Segmentation results of the proposed method on four datasets. (a) original image, (b) pre-processed, (c) binarization, (d) proposed TransUNext and (e) ground truth.}
  \label{fig:img9}
\end{figure}

Table \ref{table3} and Fig. \ref{fig:img9} present our proposed method's partial segmentation results on four datasets. Fig. \ref{fig:img9}(d) and Fig. \ref{fig:img9}(c) illustrate typical grayscale segmentation and binarization results on the four datasets. As seen from the table and figure, the segmentation result of our proposed TransUNext is very close to the ground truth, which can extract the main vessel from the background and correctly segment the vessel edge. TransUNext was shown to cope with the complexities of long and thin vessel spans and the variable morphology of the optic disc and optic cup in fundus images.

\begin{table}[!htbp]
  \centering  
  \caption{Ablation study on the DRIVE dataset.} 
  \label{table3} 
  \begin{adjustbox}{center}
    \begin{tabular}{ccccccc}
    \toprule
\textbf{Dataset}                         & \textbf{AUC}    & \textbf{SP}     & \textbf{SE}     & \textbf{Precision} & \textbf{F1-score} & \textbf{Acc}    \\
\midrule
DRIVE                           & 0.9867 & 0.9937 & 0.8643 & 0.9465    & 0.9035   & 0.9697 \\
STARE                           & 0.9869 & 0.9907 & 0.8401 & 0.8959    & 0.8671   & 0.9737 \\
CHASE-DB1                       & 0.991  & 0.9966 & 0.8817 & 0.9566    & 0.9176   & 0.9741 \\
HRF                             & 0.9887 & 0.991  & 0.8857 & 0.9098    & 0.8976   & 0.9643 \\
Baseline + all (our TransUNext) & 0.9867 & 99.85  & 93.58  & 91.28     & 18.51    & 8.55   \\
    \bottomrule
    \end{tabular}
    \end{adjustbox}
\end{table}

\subsubsection{Comparisons with SOTA methods}

\textbf{Comparison with SOTA methods on the DRIVE test dataset.} We further compare the proposed method with existing state-of-the-art (SOTA) methods to verify the effectiveness of the proposed method. The comparison in Table \ref{table4} is based on the DRIVE dataset, and TransUNext is in the leading position in terms of performance for AUC, ACC, SP, and SE. While AUC and ACC are slightly lower than \cite{10.1007/978-3-031-16434-7_51} and \cite{10.1007/978-3-031-16434-7_55}, SP and SE reach the maximum.

\begin{table}[!htbp]
  \centering  
  \caption{Comparison with SOTA methods on the DRIVE test dataset.} 
  \label{table4} 
  \begin{adjustbox}{center}
    \begin{tabular}{cccccc}
    \toprule
Methods                   & Pub.        & DRIVE   &         &        &        \\
\midrule
                          &             & AUC(\%) & ACC(\%) & SP(\%) & SE(\%) \\
\midrule
Maninis et al. \cite{10.1007/978-3-319-46723-8_17}   & MICCAI 2016 & 98.01   & 95.41   & -      & 82.8   \\
JL-UNet \cite{8341481}          & T-BME 2018  & 95.56   & \textbf{97.84}   & 98.18  & 77.92  \\
MS-NFN \cite{10.1007/978-3-030-00934-2_14}          & MICCAI 2018 & 98.07   & 95.67   & 98.19  & 78.44  \\
Yan et al. \cite{8476171}       & J-BHI 2018  & 97.5    & 95.38   & 98.2   & 76.31  \\
CE-Net \cite{8662594}           & TMI 2019    & 97.79   & 95.45   & -      & 83.09  \\
Wang et al. \cite{10.1007/978-3-030-32239-7_10}     & MICCAI 2019 & 97.72   & 95.67   & -      & 79.4   \\
DUNet \cite{JIN2019149}            & KBS 2019    & 98.02   & 95.66   & 98     & 79.63  \\
CTF-Net \cite{9098742}          & ISBI 2020   & 97.88   & 95.67   & -      & 78.49  \\
Wang et al. \cite{9119750}      & J-BHI 2020  & 98.23   & 95.81   & 98.13  & 79.91  \\
Li et al. \cite{Li_2020_WACV}        & WACV 2020   & 98.16   & 95.73   & 98.38  & 77.35  \\
CGA-Net \cite{9433813}         & ISBI 2021   & 98.65   & 96.47   & -      & 83.05  \\
SCS-Net \cite{WU2021102025}          & MIA 2021    & 98.37   & 96.97   & -      & 82.89  \\
DA-Net \cite{10.1007/978-3-031-16434-7_51}           & MICCAI 2022 & \textbf{99.03}   & \underline{97.07}   & -      & \underline{85.57}  \\
Xu et al. \cite{10.1007/978-3-031-16434-7_55}        & MICCAI 2022 & \underline{98.9}    & 97.05   & 98.28  & 84.41  \\
TransUNet (Impl) \cite{2102.04306} & -           & 97.86   & 95.32   & \underline{98.96}  & 78.39  \\
Swin-UNet (Impl) \cite{2105.05537} & -           & 97.78   & 95.4    & 98.81  & 80.05  \\
TransUNext (Impl)         & -           & 98.67   & 96.97   & \textbf{99.37}  & \textbf{86.43}  \\
    \bottomrule
    \end{tabular}
    \end{adjustbox}
    \begin{tablenotes}
     \item[1] \textbf{Note}: The best results are marked in bold and second best are marked in underline. The last three lines of data are calculated after strictly following the code implementation of the network structure in the papers.
   \end{tablenotes}
\end{table}

\textbf{Comparison with SOTA methods on the STARE test dataset.} To validate the generalization of TransUNext, Table \ref{table5} compares TransUNext to the STARE dataset, which is also at a higher level. Although the mean of our test results on the STARE dataset is weaker than \cite{9098742}, the performance on the DRIVE and CHASE-DB1 datasets is more leading.

\begin{table}[!htbp]
  \centering  
  \caption{Comparison with SOTA methods on the STARE test dataset.} 
  \label{table5} 
  \begin{adjustbox}{center}
    \begin{tabular}{cccccc}
    \toprule
Methods                   & Pub.       & STARE   &         &        &        \\
\midrule
                          &            & AUC(\%) & ACC(\%) & SP(\%) & SE(\%) \\
\midrule
JL-UNet \cite{8341481}          & T-BME 2018 & 98.01   & 96.12   & 98.46  & 75.81  \\
Yan et al. \cite{8476171}       & J-BHI 2018 & 98.33   & 96.38   & 98.57  & 77.35  \\
DUNet \cite{JIN2019149}            & KBS 2019   & 98.32   & 96.41   & 98.78  & 75.95  \\
CTF-Net \cite{9098742}          & ISBI 2020  & \textbf{99.6}    & \textbf{98.49}   & \textbf{99.34}  & \textbf{90.24}  \\
Wang et al. \cite{9119750}      & J-BHI 2020 & \underline{98.81}   & 96.73   & 98.44  & 81.86  \\
Li et al. \cite{Li_2020_WACV}        & WACV 2020  & \underline{98.81}   & 97.01   & 98.86  & 77.15  \\
SCS-Net \cite{WU2021102025}          & MIA 2021   & 98.77   & 97.36   & 98.39  & 82.07  \\
TransUNet (Impl) \cite{2102.04306} & -          & 97.74   & 95.37   & 98.67  & 73.98  \\
Swin-UNet (Impl) \cite{2105.05537} & -          & 98.32   & 96.43   & 98.64  & 79.57  \\
TransUNext (Impl)         & -          & 98.69   & \underline{97.37}   & \underline{99.07}  & \underline{84.01}  \\
    \bottomrule
    \end{tabular}
    \end{adjustbox}
\end{table}

\textbf{Comparison with SOTA methods on the CHASE-DB1 test dataset.} The comparison in Table \ref{table6} is based on the CHASE-DB1 dataset, where the SP and SE of TransUNext reach a maximum and the AUC value is slightly lower than \cite{10.1007/978-3-031-16434-7_55}.

\begin{table}[!htbp]
  \centering  
  \caption{Comparison with SOTA methods on the CHASE-DB1 test dataset.} 
  \label{table6} 
  \begin{adjustbox}{center}
    \begin{tabular}{cccccc}
    \toprule
Methods                   & Pub.        & CHASE-DB1   &         &        &        \\
\midrule
                          &             & AUC(\%) & ACC(\%) & SP(\%) & SE(\%) \\
\midrule
Maninis et al. \cite{10.1007/978-3-319-46723-8_17}   & MICCAI 2016 & 97.46     & 96.57   & -      & 76.51  \\
JL-UNet \cite{8341481}          & T-BME 2018  & 97.81     & 96.1    & 98.09  & 76.33  \\
MS-NFN \cite{10.1007/978-3-030-00934-2_14}           & MICCAI 2018 & 98.25     & 96.37   & 98.47  & 75.38  \\
Yan et al. \cite{8476171}       & J-BHI 2018  & 97.76     & 96.07   & 98.06  & 76.41  \\
CE-Net \cite{8662594}           & TMI 2019    & 98.3      & 96.89   & -      & 81.52  \\
Wang et al. \cite{10.1007/978-3-030-32239-7_10}      & MICCAI 2019 & 98.12     & 96.61   & -      & 80.74  \\
DUNet \cite{JIN2019149}            & KBS 2019    & 98.04     & 96.1    & 97.52  & 81.55  \\
CTF-Net \cite{9098742}          & ISBI 2020   & 98.47     & 96.48   & -      & 79.48  \\
Li et al. \cite{Li_2020_WACV}        & WACV 2020   & 98.51     & 96.55   & 98.23  & 79.7   \\
CGA-Net \cite{9433813}          & ISBI 2021   & 98.12     & 97.06   & -      & 86.78  \\
SCS-Net \cite{WU2021102025}          & MIA 2021    & 98.67     & 97.44   & -      & 83.65  \\
DA-Net \cite{10.1007/978-3-031-16434-7_51}           & MICCAI 2022 & 99.08     & \underline{97.66}   & -      & \underline{87.04}  \\
Xu et al. \cite{10.1007/978-3-031-16434-7_55}        & MICCAI 2022 & \textbf{99.19}     & \textbf{97.71}   & 98.55  & 85.43  \\
TransUNet (Impl) \cite{2102.04306} & -           & 97.78     & 95.4    & \underline{98.8}   & 72.04  \\
Swin-UNet (Impl) \cite{2105.05537} & -           & 98.73     & 97.38   & 98.44  & 83.61  \\
TransUNext (Impl)         & -           & \underline{99.1}      & 97.41   & \textbf{99.66}  & \textbf{88.17}  \\
    \bottomrule
    \end{tabular}
    \end{adjustbox}
\end{table}

\textbf{Comparison with SOTA methods on the HRF test dataset.} Unlike the above three datasets, the images of HRF have higher resolution, and many SOTA methods have not been validated on this dataset. Table \ref{table7} shows the results of comparing TransUNext with other methods on the HRF dataset, showing that TransUNext can accurately segment the high-resolution fundus vascular images, further demonstrating its generalization capability.

\begin{table}[!htbp]
  \centering  
  \caption{Comparison with SOTA methods on the HRF test dataset.} 
  \label{table7} 
  \begin{adjustbox}{center}
    \begin{tabular}{cccccc}
    \toprule
Methods                   & Pub.       & HRF   &         &        &        \\
\midrule
                          &            & AUC(\%) & ACC(\%) & SP(\%) & SE(\%) \\
\midrule
FullyCRF \cite{7420682}         & T-BME 2016 & -       & 91.64   & 97.09  & 59.65  \\
DRIS-GP \cite{8868109}          & TIP 2019   & -       & 94.25   & 98.98  & 66.11  \\
SCS-Net \cite{WU2021102025}          & MIA 2021   & \underline{98.42}   & \textbf{96.87}   & 98.23  & \underline{81.14}  \\
Tan et al. \cite{9740153}       & TMI 2022   & -       & 95.9    & 98.86  & 78.53  \\
TransUNet (Impl) \cite{2102.04306} & -          & 97.78   & 95.77   & \textbf{99.26}  & 65.58  \\
Swin-UNet (Impl) \cite{2105.05537} & -          & 97.81   & 95.83   & 98.85  & 80.99  \\
TransUNext (Impl)         & -          & \textbf{98.87}   & \underline{96.43}   & \underline{99.1}   & \textbf{88.57}  \\
    \bottomrule
    \end{tabular}
    \end{adjustbox}
\end{table}

\begin{figure}
  \centering
  \includegraphics[width=1\textwidth]{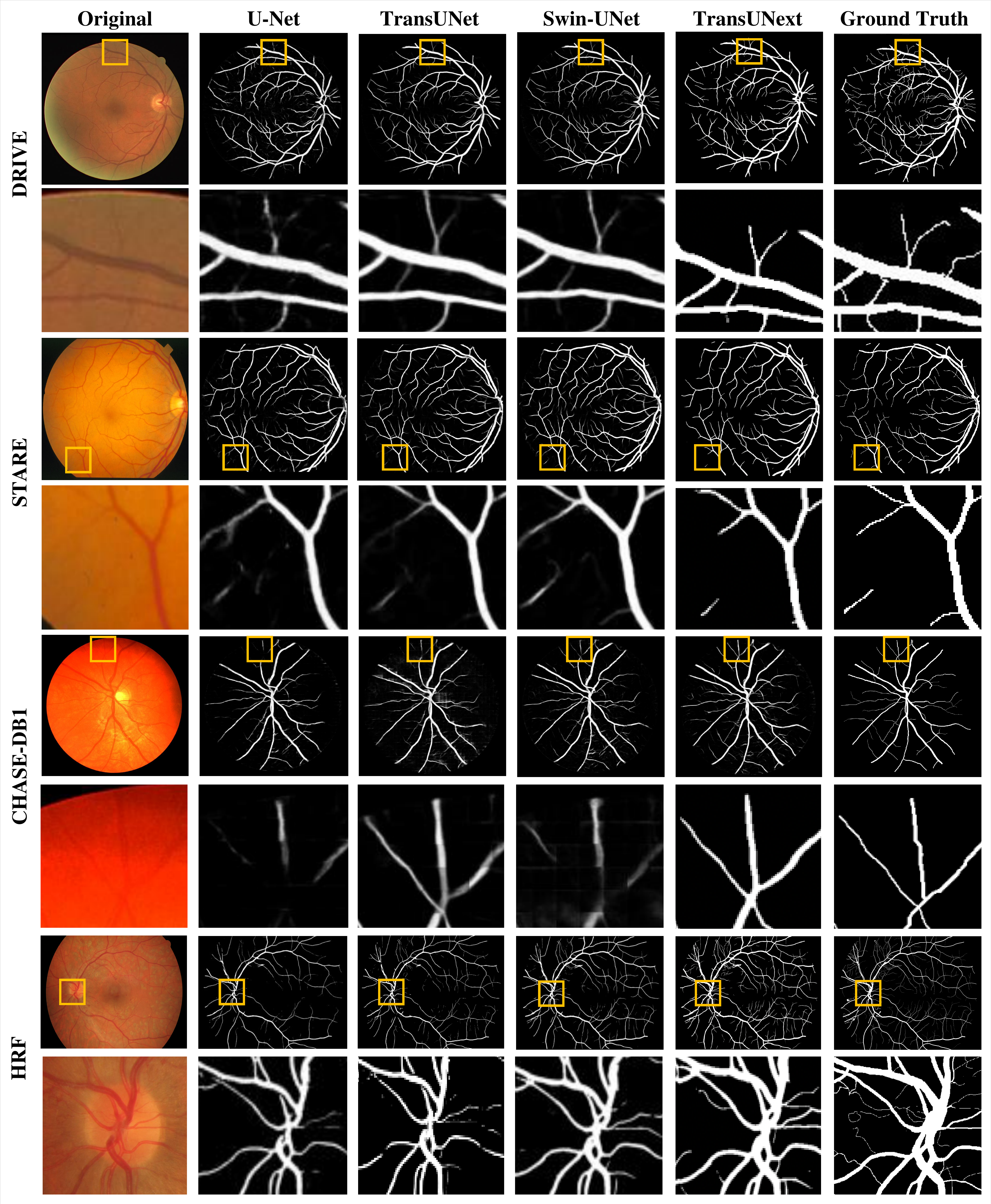}
  \caption{Comparative results of the SOTA methods and our proposed TransUNext on four datasets, where the second, fourth, sixth and eighth rows give the local zoomed-in results of the vessel ends and at the optic disc.}
  \label{fig:img10}
\end{figure}

\textbf{Qualitative visualizations of the proposed TransUNext and other SOTA methods.} Fig. \ref{fig:img10} compares the visualization results of several SOTA methods on four datasets. Our method achieves satisfactory segmentation results from the locally zoom-in images. Compared with other improved approaches, such as TransUNet \cite{2102.04306} and Swin-UNet \cite{2105.05537}, our proposed TransUNext can detect retinal vessels more correctly and reduce misclassified retinal vessel pixels. In addition, better recognition is achieved for tiny vessels and edge regions.

\subsubsection{Discussion}
In TransUNext, the TransNeXt Block is used to improve the base block of traditional U-Net, while combining the advantages of Transformer and ConvNeXt, avoiding the massive pre-training of Transformer and with less computational overhead. In addition, the GMSF mechanism replaces the skip connection of U-Net for semantic and spatial global multi-scale fusion, which allows the overall architecture of TransUNext to follow the design of Fig. \ref{fig:img1}(e). Extensive experiments have demonstrated that TransUNext brings significant performance gains on retinal vessel segmentation tasks. In the future, we will further improve the hybrid approach of Transformer and ConvNeXt modules, such as the parallel hybrid architecture design of both, and explore how the Transformer can be pre-trained end-to-end in medical images to improve the segmentation performance further.

\subsection{Conclusion}
We propose a retinal vessel segmentation method, TransUNext, that integrates an Efficient Self-attention Mechanism into the encoder and decoder of U-Net to capture both local features and global dependencies with minimal computational overhead. Unlike other hybrid architectures that embed modular Transformers into CNNs, TransUNext constructs hybrid modules consisting of Transformer and ConvNeXt throughout the model, which continuously aggregates representations from global and local features to give full play to the superiority of both. Among them, the GMSF module upgrades the skip connection, fuses high-level semantic and low-level detail information and eliminates the high and low-level semantic differences. In addition, we design the TransNeXt Block as the base block of the U-shaped architecture, which not only captures local features but also avoids the information loss caused by the compressed dimension when the information is converted between the feature spaces of different dimensions.

Our TransUNext significantly outperforms other SOTA methods for retinal vessel segmentation on DRIVE, SATRE, CHASE-DB1, and HRF datasets and can be applied to other high-resolution or strip objects and lesions segmentation tasks potentially.

\textbf{Ethical approval}

This article does not contain any studies with human participants or animals performed by any of the authors.

\textbf{Competing Interests}

The authors declare that they have no conflict of interest.

\bibliographystyle{splncs04}
\bibliography{citation}

\begin{thebibliography}{10}
\providecommand{\url}[1]{\texttt{#1}}
\providecommand{\urlprefix}{URL }
\providecommand{\doi}[1]{https://doi.org/#1}

\bibitem{2105.05537}
Cao, H., Wang, Y., Chen, J., Jiang, D., Zhang, X., Tian, Q., Wang, M.: Swin-unet: Unet-like pure transformer for medical image segmentation (2021)

\bibitem{34715}
Chaudhuri, S., Chatterjee, S., Katz, N., Nelson, M., Goldbaum, M.: Detection of blood vessels in retinal images using two-dimensional matched filters. IEEE Transactions on Medical Imaging  \textbf{8}(3),  263--269 (1989). \doi{10.1109/42.34715}

\bibitem{2102.04306}
Chen, J., Lu, Y., Yu, Q., Luo, X., Adeli, E., Wang, Y., Lu, L., Yuille, A.L., Zhou, Y.: Transunet: Transformers make strong encoders for medical image segmentation (2021)

\bibitem{8868109}
Cherukuri, V., Kumar~B.G., V., Bala, R., Monga, V.: Deep retinal image segmentation with regularization under geometric priors. IEEE Transactions on Image Processing  \textbf{29},  2552--2567 (2020). \doi{10.1109/TIP.2019.2946078}

\bibitem{2010.11929}
Dosovitskiy, A., Beyer, L., Kolesnikov, A., Weissenborn, D., Zhai, X., Unterthiner, T., Dehghani, M., Minderer, M., Heigold, G., Gelly, S., Uszkoreit, J., Houlsby, N.: An image is worth 16x16 words: Transformers for image recognition at scale (2020)

\bibitem{6224174}
Fraz, M.M., Remagnino, P., Hoppe, A., Uyyanonvara, B., Rudnicka, A.R., Owen, C.G., Barman, S.A.: An ensemble classification-based approach applied to retinal blood vessel segmentation. IEEE Transactions on Biomedical Engineering  \textbf{59}(9),  2538--2548 (2012). \doi{10.1109/TBME.2012.2205687}

\bibitem{6019055}
Gegundez-Arias, M.E., Aquino, A., Bravo, J.M., Marin, D.: A function for quality evaluation of retinal vessel segmentations. IEEE Transactions on Medical Imaging  \textbf{31}(2),  231--239 (2012). \doi{10.1109/TMI.2011.2167982}

\bibitem{8662594}
Gu, Z., Cheng, J., Fu, H., Zhou, K., Hao, H., Zhao, Y., Zhang, T., Gao, S., Liu, J.: Ce-net: Context encoder network for 2d medical image segmentation. IEEE Transactions on Medical Imaging  \textbf{38}(10),  2281--2292 (2019). \doi{10.1109/TMI.2019.2903562}

\bibitem{Hatamizadeh_2022_WACV}
Hatamizadeh, A., Tang, Y., Nath, V., Yang, D., Myronenko, A., Landman, B., Roth, H.R., Xu, D.: Unetr: Transformers for 3d medical image segmentation. In: Proceedings of the IEEE/CVF Winter Conference on Applications of Computer Vision (WACV). pp. 574--584 (January 2022)

\bibitem{1704.04861}
Howard, A.G., Zhu, M., Chen, B., Kalenichenko, D., Wang, W., Weyand, T., Andreetto, M., Adam, H.: Mobilenets: Efficient convolutional neural networks for mobile vision applications (2017)

\bibitem{JIN2019149}
Jin, Q., Meng, Z., Pham, T.D., Chen, Q., Wei, L., Su, R.: Dunet: A deformable network for retinal vessel segmentation. Knowledge-Based Systems  \textbf{178},  149--162 (2019). \doi{https://doi.org/10.1016/j.knosys.2019.04.025}, \url{https://www.sciencedirect.com/science/article/pii/S0950705119301984}

\bibitem{Li_2020_WACV}
Li, L., Verma, M., Nakashima, Y., Nagahara, H., Kawasaki, R.: Iternet: Retinal image segmentation utilizing structural redundancy in vessel networks. In: Proceedings of the IEEE/CVF Winter Conference on Applications of Computer Vision (WACV) (March 2020)

\bibitem{LI20127600}
Li, Q., You, J., Zhang, D.: Vessel segmentation and width estimation in retinal images using multiscale production of matched filter responses. Expert Systems with Applications  \textbf{39}(9),  7600--7610 (2012). \doi{https://doi.org/10.1016/j.eswa.2011.12.046}, \url{https://www.sciencedirect.com/science/article/pii/S0957417411017179}

\bibitem{10.1007/978-3-031-16443-9_23}
Liu, W., Tian, T., Xu, W., Yang, H., Pan, X., Yan, S., Wang, L.: Phtrans: Parallelly aggregating global and local representations for medical image segmentation. In: Wang, L., Dou, Q., Fletcher, P.T., Speidel, S., Li, S. (eds.) Medical Image Computing and Computer Assisted Intervention -- MICCAI 2022. pp. 235--244. Springer Nature Switzerland, Cham (2022)

\bibitem{Liu_2021_ICCV}
Liu, Z., Lin, Y., Cao, Y., Hu, H., Wei, Y., Zhang, Z., Lin, S., Guo, B.: Swin transformer: Hierarchical vision transformer using shifted windows. In: Proceedings of the IEEE/CVF International Conference on Computer Vision (ICCV). pp. 10012--10022 (October 2021)

\bibitem{Liu_2022_CVPR}
Liu, Z., Mao, H., Wu, C.Y., Feichtenhofer, C., Darrell, T., Xie, S.: A convnet for the 2020s. In: Proceedings of the IEEE/CVF Conference on Computer Vision and Pattern Recognition (CVPR). pp. 11976--11986 (June 2022)

\bibitem{Luo_2020}
Luo, Z., Jia, Y.: The comparison of retinal vessel segmentation methods in fundus images. Journal of Physics: Conference Series  \textbf{1574}(1),  012160 (jun 2020). \doi{10.1088/1742-6596/1574/1/012160}, \url{https://dx.doi.org/10.1088/1742-6596/1574/1/012160}

\bibitem{10.1007/978-3-319-46723-8_17}
Maninis, K.K., Pont-Tuset, J., Arbel{\'a}ez, P., Van~Gool, L.: Deep retinal image understanding. In: Ourselin, S., Joskowicz, L., Sabuncu, M.R., Unal, G., Wells, W. (eds.) Medical Image Computing and Computer-Assisted Intervention -- MICCAI 2016. pp. 140--148. Springer International Publishing, Cham (2016)

\bibitem{7420682}
Orlando, J.I., Prokofyeva, E., Blaschko, M.B.: A discriminatively trained fully connected conditional random field model for blood vessel segmentation in fundus images. IEEE Transactions on Biomedical Engineering  \textbf{64}(1),  16--27 (2017). \doi{10.1109/TBME.2016.2535311}

\bibitem{4336179}
Ricci, E., Perfetti, R.: Retinal blood vessel segmentation using line operators and support vector classification. IEEE Transactions on Medical Imaging  \textbf{26}(10),  1357--1365 (2007). \doi{10.1109/TMI.2007.898551}

\bibitem{10.1007/978-3-319-24574-4_28}
Ronneberger, O., Fischer, P., Brox, T.: U-net: Convolutional networks for biomedical image segmentation. In: Navab, N., Hornegger, J., Wells, W.M., Frangi, A.F. (eds.) Medical Image Computing and Computer-Assisted Intervention -- MICCAI 2015. pp. 234--241. Springer International Publishing, Cham (2015)

\bibitem{1677727}
Soares, J., Leandro, J., Cesar, R., Jelinek, H., Cree, M.: Retinal vessel segmentation using the 2-d gabor wavelet and supervised classification. IEEE Transactions on Medical Imaging  \textbf{25}(9),  1214--1222 (2006). \doi{10.1109/TMI.2006.879967}

\bibitem{1282003}
Staal, J., Abramoff, M., Niemeijer, M., Viergever, M., van Ginneken, B.: Ridge-based vessel segmentation in color images of the retina. IEEE Transactions on Medical Imaging  \textbf{23}(4),  501--509 (2004). \doi{10.1109/TMI.2004.825627}

\bibitem{9740153}
Tan, Y., Yang, K.F., Zhao, S.X., Li, Y.J.: Retinal vessel segmentation with skeletal prior and contrastive loss. IEEE Transactions on Medical Imaging  \textbf{41}(9),  2238--2251 (2022). \doi{10.1109/TMI.2022.3161681}

\bibitem{NIPS2017_3f5ee243}
Vaswani, A., Shazeer, N., Parmar, N., Uszkoreit, J., Jones, L., Gomez, A.N., Kaiser, L.u., Polosukhin, I.: Attention is all you need. In: Guyon, I., Luxburg, U.V., Bengio, S., Wallach, H., Fergus, R., Vishwanathan, S., Garnett, R. (eds.) Advances in Neural Information Processing Systems. vol.~30. Curran Associates, Inc. (2017), \url{https://proceedings.neurips.cc/paper/2017/file/3f5ee243547dee91fbd053c1c4a845aa-Paper.pdf}

\bibitem{10.1007/978-3-030-32239-7_10}
Wang, B., Qiu, S., He, H.: Dual encoding u-net for retinal vessel segmentation. In: Shen, D., Liu, T., Peters, T.M., Staib, L.H., Essert, C., Zhou, S., Yap, P.T., Khan, A. (eds.) Medical Image Computing and Computer Assisted Intervention -- MICCAI 2019. pp. 84--92. Springer International Publishing, Cham (2019)

\bibitem{10.1007/978-3-031-16434-7_51}
Wang, C., Xu, R., Xu, S., Meng, W., Zhang, X.: Da-net: Dual branch transformer and adaptive strip upsampling for retinal vessels segmentation. In: Wang, L., Dou, Q., Fletcher, P.T., Speidel, S., Li, S. (eds.) Medical Image Computing and Computer Assisted Intervention -- MICCAI 2022. pp. 528--538. Springer Nature Switzerland, Cham (2022)

\bibitem{9433813}
Wang, C., Xu, R., Zhang, Y., Xu, S., Zhang, X.: Retinal vessel segmentation via context guide attention net with joint hard sample mining strategy. In: 2021 IEEE 18th International Symposium on Biomedical Imaging (ISBI). pp. 1319--1323 (2021). \doi{10.1109/ISBI48211.2021.9433813}

\bibitem{9119750}
Wang, D., Haytham, A., Pottenburgh, J., Saeedi, O., Tao, Y.: Hard attention net for automatic retinal vessel segmentation. IEEE Journal of Biomedical and Health Informatics  \textbf{24}(12),  3384--3396 (2020). \doi{10.1109/JBHI.2020.3002985}

\bibitem{Wang2021}
Wang, J., Li, X., Lv, P., Shi, C.: Serr-u-net: Squeeze-and-excitation residual and recurrent block-based u-net for automatic vessel segmentation in retinal image. Computational and Mathematical Methods in Medicine  \textbf{2021},  5976097 (Aug 2021). \doi{10.1155/2021/5976097}, \url{https://doi.org/10.1155/2021/5976097}

\bibitem{9098742}
Wang, K., Zhang, X., Huang, S., Wang, Q., Chen, F.: Ctf-net: Retinal vessel segmentation via deep coarse-to-fine supervision network. In: 2020 IEEE 17th International Symposium on Biomedical Imaging (ISBI). pp. 1237--1241 (2020). \doi{10.1109/ISBI45749.2020.9098742}

\bibitem{2006.04768}
Wang, S., Li, B.Z., Khabsa, M., Fang, H., Ma, H.: Linformer: Self-attention with linear complexity (2020)

\bibitem{10.1007/978-3-030-87193-2_11}
Wang, W., Chen, C., Ding, M., Yu, H., Zha, S., Li, J.: Transbts: Multimodal brain tumor segmentation using transformer. In: de~Bruijne, M., Cattin, P.C., Cotin, S., Padoy, N., Speidel, S., Zheng, Y., Essert, C. (eds.) Medical Image Computing and Computer Assisted Intervention -- MICCAI 2021. pp. 109--119. Springer International Publishing, Cham (2021)

\bibitem{Wang_2018_ECCV_Workshops}
Wang, X., Yu, K., Wu, S., Gu, J., Liu, Y., Dong, C., Qiao, Y., Change~Loy, C.: Esrgan: Enhanced super-resolution generative adversarial networks. In: Proceedings of the European Conference on Computer Vision (ECCV) Workshops (September 2018)

\bibitem{10.1145/3348416.3348425}
Wang, Z., Lin, J., Wang, R., Zheng, W.: Data augmentation is more important than model architectures for retinal vessel segmentation. In: Proceedings of the 2019 International Conference on Intelligent Medicine and Health. p. 48–52. ICIMH 2019, Association for Computing Machinery, New York, NY, USA (2019). \doi{10.1145/3348416.3348425}, \url{https://doi.org/10.1145/3348416.3348425}

\bibitem{WU2021102025}
Wu, H., Wang, W., Zhong, J., Lei, B., Wen, Z., Qin, J.: Scs-net: A scale and context sensitive network for retinal vessel segmentation. Medical Image Analysis  \textbf{70},  102025 (2021). \doi{https://doi.org/10.1016/j.media.2021.102025}, \url{https://www.sciencedirect.com/science/article/pii/S1361841521000712}

\bibitem{10.1007/978-3-030-00934-2_14}
Wu, Y., Xia, Y., Song, Y., Zhang, Y., Cai, W.: Multiscale network followed network model for retinal vessel segmentation. In: Frangi, A.F., Schnabel, J.A., Davatzikos, C., Alberola-L{\'o}pez, C., Fichtinger, G. (eds.) Medical Image Computing and Computer Assisted Intervention -- MICCAI 2018. pp. 119--126. Springer International Publishing, Cham (2018)

\bibitem{Xie_2017_CVPR}
Xie, S., Girshick, R., Dollar, P., Tu, Z., He, K.: Aggregated residual transformations for deep neural networks. In: Proceedings of the IEEE Conference on Computer Vision and Pattern Recognition (CVPR) (July 2017)

\bibitem{10.1007/978-3-030-87199-4_16}
Xie, Y., Zhang, J., Shen, C., Xia, Y.: Cotr: Efficiently bridging cnn and transformer for 3d medical image segmentation. In: de~Bruijne, M., Cattin, P.C., Cotin, S., Padoy, N., Speidel, S., Zheng, Y., Essert, C. (eds.) Medical Image Computing and Computer Assisted Intervention -- MICCAI 2021. pp. 171--180. Springer International Publishing, Cham (2021)

\bibitem{10.1007/978-3-031-16434-7_55}
Xu, R., Zhao, J., Ye, X., Wu, P., Wang, Z., Li, H., Chen, Y.W.: Local-region and cross-dataset contrastive learning for retinal vessel segmentation. In: Wang, L., Dou, Q., Fletcher, P.T., Speidel, S., Li, S. (eds.) Medical Image Computing and Computer Assisted Intervention -- MICCAI 2022. pp. 571--581. Springer Nature Switzerland, Cham (2022)

\bibitem{8341481}
Yan, Z., Yang, X., Cheng, K.T.: Joint segment-level and pixel-wise losses for deep learning based retinal vessel segmentation. IEEE Transactions on Biomedical Engineering  \textbf{65}(9),  1912--1923 (2018). \doi{10.1109/TBME.2018.2828137}

\bibitem{8476171}
Yan, Z., Yang, X., Cheng, K.T.: A three-stage deep learning model for accurate retinal vessel segmentation. IEEE Journal of Biomedical and Health Informatics  \textbf{23}(4),  1427--1436 (2019). \doi{10.1109/JBHI.2018.2872813}

\bibitem{10.5555/2975865.2975874}
Yang, Y., Huang, S., Rao, N.: An automatic hybrid method for retinal blood vessel extraction. Int. J. Appl. Math. Comput. Sci.  \textbf{18}(3),  399–407 (sep 2008)

\bibitem{7055281}
Zhao, Y., Rada, L., Chen, K., Harding, S.P., Zheng, Y.: Automated vessel segmentation using infinite perimeter active contour model with hybrid region information with application to retinal images. IEEE Transactions on Medical Imaging  \textbf{34}(9),  1797--1807 (2015). \doi{10.1109/TMI.2015.2409024}

\bibitem{2109.03201}
Zhou, H.Y., Guo, J., Zhang, Y., Yu, L., Wang, L., Yu, Y.: nnformer: Interleaved transformer for volumetric segmentation (2021)

\end{thebibliography}
\end{document}